\documentclass[11pt]{article}
\pdfoutput=1
\usepackage{graphicx}
\usepackage[margin=1in]{geometry}
\usepackage{caption}
\usepackage{subcaption}
\usepackage{amsmath}
\begin{document}

\begin{titlepage}
\begin{center}

{ \huge \bfseries Primordial black holes in non-Gaussian regimes}\\[1cm]

Sam Young, Christian T.~Byrnes\\[0.5cm]
Department of Physics and Astronomy, Pevensey II Building, University of Sussex, BN1 9RH, UK\\[1cm]

\today\\[1cm]

\end{center}

Primordial black holes (PBHs) can form in the early Universe from the collapse of rare, large density fluctuations. They have never been observed, but this fact is enough to constrain the amplitude of fluctuations on very small scales which cannot be otherwise probed. Because PBHs form only in very rare large fluctuations, the number of PBHs formed is extremely sensitive to changes in the shape of the tail of the fluctuation distribution - which depends on the amount of non-Gaussianity present. We first study how local non-Gaussianity of arbitrary size up to fifth order affects the abundance and constraints from PBHs, finding that they depend strongly on even small amounts of non-Gaussianity and the upper bound on the allowed amplitude of the power spectrum can vary by several orders of magnitude. The sign of the non-linearity parameters ($f_{NL}$, $g_{NL}$, etc) are particularly important. We also study the abundance and constraints from PBHs in the curvaton scenario, in which case the complete non-linear probability distribution is known, and find that truncating to any given order (i.e.~to order $f_{NL}$ or $g_{NL}$, etc) does not give accurate results.

\end{titlepage}

\section{Introduction}

Primordial black holes (PBHs) have historically been used to study the small scales of the primordial universe. Whilst they have never been detected, this fact is enough to rule out or at least constrain many different cosmological models (see Refs.~\cite{D,E,F,G,O}). Theoretical arguments suggest that PBHs can form from the collapse of large density perturbations during radiation domination \cite{AG}. If the density perturbation at horizon crossing exceeds a threshold value, then gravity will overcome pressure forces and that region collapses to form a PBH with mass of order the horizon mass.

There are tight observational constraints on the abundance of PBHs. These constraints come from their gravitational effects and results of the Hawking radiation from their evaporation. For recent updates and a compilation of the constraints see Refs \cite{B,C,W}. The various constraints place an upper limit on the mass fraction of the Universe contained within PBHs at the time of formation, $\beta$. The constraints vary from $\beta=10^{-27}$ to $\beta=10^{-5}$. These constraints can be used to constrain the primordial power spectrum on small scales, and hence models of inflation. Since PBHs form from the rare, large fluctuations in the extreme tail of the probability distribution function (PDF), any non-Gaussianity can significantly affect the number of PBHs formed. PBH formation can therefore be used to probe both the amplitude and non-Gaussianity of the primordial fluctuations on small scales.

In order for a significant number of PBHs to form, the power spectrum on small scales needs to be of order $10^{-2}$, orders of magnitude larger than on cosmic scales. Although a spectral index smaller than 1 has recently been observed by Planck, indicating a red spectrum, it is possible that the running of the spectral index turns up on smaller scales, and produces a lot of power at such scales. This is possible in models such as the running-mass model, the inflating curvaton and hybrid inflation \cite{A, R, S, T, U, V, AI,Linde:2012bt,Lin:2012gs}. Other possibilities include peaks in the power spectrum\cite{W} or a phase transition after inflation \cite{Barrow:1992hq}.

The effects of non-Gaussianity on PBH formation were first studied by Bullock and Primack \cite{H}, and Ivanov \cite{I} - reaching opposite conclusions on whether non-Gaussianity enhances or suppresses the number of PBHs formed. Lyth \cite{AL} studied the constraints from PBH formation on the primordial curvature perturbation for cases where it has the form $\zeta=\pm\left(x^{2}-\langle x^{2}\rangle\right)$, where $x$ has a Gaussian distribution. The minus sign can be expected from the linear era of the hybrid inflation waterfall, where the positive sign might arise if $\zeta$ is generated after inflation by a curvaton-type mechanism. More recently, the effects of non-Gaussianity have been studied by Byrnes et al \cite{AK}, who studied the effects of quadratic and cubic non-Gaussianity in the local model of non-Gaussianity, and Shandera et al \cite{AF}, who considered small deviations from a Gaussian distribution, finding that whether PBH formation is enhanced or suppressed depends on the type of non-Gaussianity. The effects of non-Gaussianity in the curvaton model have also been studied recently by Bugaev and Klimai \cite{AH, AJ}, who calculated constraints and PBH mass spectra for a chi-squared distribution. Seery and Hidalgo \cite{AM} showed how to obtain the probability distribution of the curvature perturbation working directly from the $n$-point correlation functions (which come from quantum field theory calculations) and discussed the possibility of using the constraints of PBHs to discriminate between models of inflation.

In this paper, we will go beyond earlier work and calculate the effects of arbitrarily large non-Gaussianity in the local model to 5th order, including terms of each type simultaneously. We also consider the curvaton model where a full non-linear solution for the curvature perturbation is available in the \emph{sudden decay approximation} \cite{AD}. It is found in this case that using a perturbative approach by deriving the non-Gaussianity parameters ($f_{NL}$, $g_{NL}$, etc) and using the local model of non-Gaussianity disagrees strongly with the full solution - and so care needs to be taken when performing these calculations.

In Section 2, we review the calculation of the PBH abundance constraints in the standard Gaussian case. In Section 3, we review the work completed by Byrnes et al \cite{AK} calculating the effects of quadratic and cubic non-Gaussianity in the local model, before extending this to higher orders. The expert reader may skip to Sec.~\ref{sec:higher-order}. In Section 4 we discuss the effects of a hierarchical scaling between the non-Gaussianity parameters ($g_{NL}\propto f_{NL}^{2}$, $h_{NL}\propto f_{NL}^{3}$, etc), and in Section 5 we calculate the constraints on the primordial power spectrum in the curvaton model. We conclude with a summary in Section 6.

\section{PBHs in a Gaussian universe}

Whilst the condition required for collapse to form a PBH has traditionally been stated in terms of the smoothed density contrast at horizon crossing, $\delta_{hor}(R)$, we will follow Ref.~\cite{AK} and work with the curvature perturbation, $\zeta$. PBHs form in regions where the curvature perturbation is greater than a critical value, $\zeta_c\simeq0.7-1.2$ \cite{Green:2004wb}. There is some uncertainty on the exact critical value, and it has a dependence upon the profile of the over density \cite{Shibata:1999zs,I,AN}. For simplicity, we will usually take $\zeta_c=1$, it would be straightforward to choose any other value if required.
It was initially thought that there was an upper limit on the amplitude of the fluctuation that would form a PBH, with larger fluctuations forming a separate universe, however, this has been shown not to be the case \cite{P}. Integrating over the fluctuations which form PBHs, the initial PBH mass fraction of the Universe is:
\begin{equation}
\label{pressschecter}
\beta\equiv\frac{\rho_{PBH}}{\rho_{total}}{\Big|}_{formation}\simeq\int^{\infty}_{\zeta_{c}}P\left(\zeta\right)d\zeta,
\end{equation}
where $\zeta_{c}$ is the critical value for PBH production and $P(\zeta)$ is the probability distribution function. The above equation is not exact, for example due to the uncertainty in the fraction of mass within a horizon sized patch (whose average density is above the critical one) which will collapse to form a black hole. This is related to uncertainty of the overdensity profile and the critical value required for collapse, see e.g.~\cite{Niemeyer:1999ak,Yokoyama:1998xd,Hawke:2002rf,Musco:2012au} and references therein. Fortunately a numerical factor of order unity leads to only a small uncertainty in the constraints on $\sigma$ due to the logarithm, see  Eq.~(\ref{sigmaconstraint}). Order unity non-linearity parameters are much more important than a numerical coefficient multiplying the integral in (\ref{pressschecter}). For Gaussian fluctuations:
\begin{equation}
\label{gaussian}
P(\zeta)=\frac{1}{\sqrt{2\pi}\sigma}\exp\left(-\frac{1}{2}\frac{\zeta^{2}}{\sigma^{2}}\right),
\end{equation}
and so:
\begin{equation}
\label{gaussianintegral}
\beta\simeq\frac{1}{\sqrt{2\pi}\sigma}\int^{\infty}_{\zeta_{c}}\exp\left(-\frac{1}{2}\frac{\zeta^{2}}{\sigma^{2}}\right)d\zeta=\frac{1}{2} \textrm{erfc}\left(\frac{\zeta_{c}}{\sqrt{2}\sigma}\right).
\end{equation}
Because PBHs form in extremely rare large fluctuations in the tail of the probability distribution, one can use the large $x$ limit of $\textrm{erfc}(x)$ and show that \cite{AK}:
\begin{equation}
\frac{\sigma}{\zeta_{c}}\simeq\sigma={\cal P}_{\zeta}^{1/2}\simeq\sqrt{\frac{1}{2\ln\left(\frac{1}{\beta}\right)}}.
\label{sigmaconstraint}
\end{equation}
Note that $\sigma$ depends only logarithmically on $\beta$, this remains true once the effects of non-Gaussianity are taken into account. Taking $\zeta_{c}=1$, for $\beta=10^{-20}$ we obtain $\sigma=0.11$ and for $\beta=10^{-5}$ we obtain $\sigma=0.23$.

The variance of the probability distribution is related to the power spectrum of the curvature perturbation by $\sigma^{2}\approx {\cal P}_{\zeta}$. The constraints obtained in this manner differ by ${\cal O}(10\%)$ to those obtained from a full Press-Schechter calculation which includes a window function to smooth the curvature perturbation, as performed in \cite{Matarrese:2000iz,AC}. For $\beta=10^{-20}$ the full calculation gives ${\cal P}_{\zeta}^{1/2}=0.12$ \cite{B}, as opposed to ${\cal P}_{\zeta}^{1/2}=0.11$ obtained with Eq.~(\ref{sigmaconstraint}). In the case of chi-squared non-Gaussianity, a calculation used the smoothed pdf has also been performed \cite{O} and gives reasonable agreement with the approach we use here.

\section{PBHs and local non-Gaussianity}

We consider the effects of non-Gaussianity in the local model on the abundance of PBHs and the constraints we can place on the power spectrum. We will first review the work completed by Byrnes et al \cite{AK} and discuss the effects of quadratic and cubic local non-Gaussianity, before moving onto the effects of higher order terms in Sec.~\ref{sec:higher-order}.

\subsection{Quadratic non-Gaussianity}

\begin{figure}[t]
\centering
\begin{subfigure}{0.5\textwidth}
	\centering
	\includegraphics[width=\linewidth]{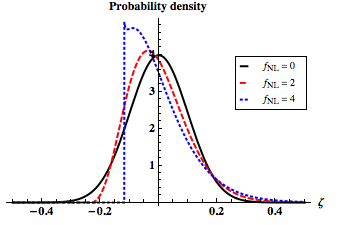}
\end{subfigure}%
\begin{subfigure}{0.5\textwidth}
	\centering
	\includegraphics[width=\linewidth]{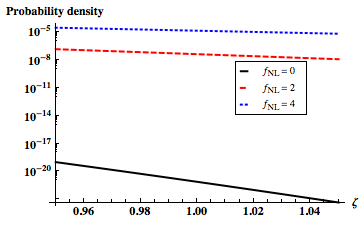}
\end{subfigure}
 \caption{The left plot shows the effect of positive $f_{NL}$ on the PDF. For negative $f_{NL}$ the PDFs are simply reflected in the y-axis. We see that the $f_{NL}$ term skews the distribution. The right plot shows the tail of the PDF where PBHs form - note that this is a logarithmic plot of the PDF. A relatively small change in $f_{NL}$ has a large effect on the number of PBHs produced - by many orders of magnitude. For these plots, we have taken $\sigma=0.1$.}
\label{fnlpdfs}
\end{figure}

We take the model of local non-Gaussianity to be
\begin{equation}
\label{quadraticNonGaussianity}
\zeta=\zeta_{g}+\frac{3}{5}f_{NL}\left(\zeta_{g}^{2}-\sigma^{2}\right).
\end{equation}
The $\sigma^{2}$ term is included to ensure that the expectation value for the curvature perturbation remains zero, $\langle\zeta\rangle=0$. Solving this equation to find $\zeta_{g}$ as a function of $\zeta$ gives two solutions
\begin{equation}
\zeta_{g\pm}(\zeta)=\frac{5}{6f_{NL}}\left[ -1 \pm \sqrt{1+\frac{12f_{NL}}{5} \left( \frac{3f_{NL}\sigma^{2}}{5}+\zeta\right)} \right].
\end{equation}
We can make a formal change of variable using
\begin{equation}
P_{NG}(\zeta)d\zeta=\sum_{i=1}^{n}\left|\frac{d\zeta_{g,i}(\zeta)}{d\zeta}\right|P_{G}\left(\zeta_{g,i}(\zeta)\right)d\zeta,
\label{changeofvariable}
\end{equation}
where $i$ is the sum over all solutions, to find the non-Gaussian probability distribution function (PDF). The non-Gaussian distribution is then given by:
\begin{equation}
\label{quadraticpdfeqn}
P_{NG}(\zeta)d\zeta=\frac{d\zeta}{\sqrt{2\pi}\sigma\sqrt{1+\frac{12f_{NL}}{5}\left(\frac{3f_{NL}\sigma^{2}}{5}+\zeta\right)}}\left(\epsilon_{+}+\epsilon_{-}\right),
\end{equation}
where
\begin{equation}
\epsilon_{\pm}=\exp\left(-\frac{\zeta_{g\pm}(\zeta)^{2}}{2\sigma^{2}}\right),
\end{equation}
and the initial PBH mass fraction is given by
\begin{equation}
\beta\simeq\int_{\zeta_{c}}^{\zeta_{max}}P_{NG}(\zeta)d\zeta.
\end{equation}
If $f_{NL}$ is positive (or zero) then $\zeta_{max}=\infty$, but if $f_{NL}$ is negative then $\zeta$ is bound from above and $\zeta_{max}$ is given by
\begin{equation}
\zeta_{max}=-\frac{5}{12f_{NL}}\left(1+\frac{36f_{NL}^{2}\sigma^{2}}{25}\right).
\end{equation}
Figure \ref{fnlpdfs} shows the effect of $f_{NL}$ on the probability density function. The primary effect of $f_{NL}$ is to skew the distribution - for positive $f_{NL}$ we see a peak for negative values of $\zeta$, with a large tail for positive values (and vice versa for negative $f_{NL}$). The right panel shows a log plot of the effect of positive $f_{NL}$ on the tail of the PDF where PBH formation occurs. We see that, for positive $f_{NL}$, as $f_{NL}$ is increased the amplitude of the large tail increases dramatically. For negative values of $f_{NL}$, $\zeta$ is bounded from above, $\zeta<1$, and we would see no PBH formation for these values (by increasing $\sigma$ significantly, one can form PBHs for significantly negative $f_{NL}$, although we will see later that unless remarkable fine tuning occurs, this leads to an overproduction of PBHs).

We now use the observational constraints on $\beta$ to place constraints on the power spectrum. This is most easily calculated by making a transformation to a new variable $y$:
\begin{equation}
\label{y}
y=\frac{\zeta_{g\pm}(\zeta)}{\sigma},
\end{equation}
which has unit variance. For $f_{NL}>0$ we have
\begin{equation}
\beta\simeq\frac{1}{\sqrt{2\pi}}\left(\int^{\infty}_{y_{c+}}e^{-\frac{y^{2}}{2}}dy+\int^{y_{c-}}_{-\infty}e^{-\frac{y^{2}}{2}}dy\right),
\end{equation}
and for $f_{NL}<0$
\begin{equation}
\beta\simeq\frac{1}{\sqrt{2\pi}}\int^{y_{c+}}_{y_{c-}}e^{-\frac{y^{2}}{2}}dy,
\end{equation}
where $y_{c\pm}$ are the values of y corresponding to the threshold for PBH formation, $\zeta_{c}$:
\begin{equation}
y_{c\pm}=\frac{\zeta_{g\pm}(\zeta_{c})}{\sigma}.
\end{equation}
The expression for $\beta$ is then solved numerically using the tight and weak constraints, $\beta=10^{-20}$ and $10^{-5}$ respectively, to find a value for $\sigma$. The variance of $\zeta$ is then given by \cite{Boubekeur:2005fj,Y}
\begin{equation}
{\cal P}_{\zeta}=\sigma^{2}+4\left(\frac{3f_{NL}}{5}\right)^2\sigma^{4}\ln(kL),
\end{equation}
where the cut-off scale $L\approx1/H$ is of order the horizon scale, $k$ is the scale of interest and $\ln(kL)$ is typically ${\cal O}(1)$ (treating it as exactly 1 leads to percent level corrections, provided that $\sigma$ is small - we have numerically checked this).

\begin{figure}[t]
 \centerline{\includegraphics[scale=0.8]{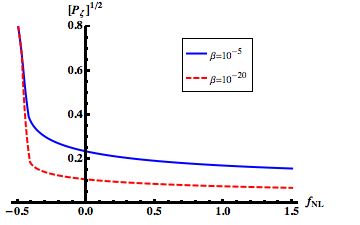}}
 \caption{This plot shows how the constraints on the square root of the power spectrum due to PBHs depend on $f_{NL}$. The constraints for 2 values of $\beta$ are shown - note that, although $\beta$ changes by 15 orders of magnitude, the constraints only change by a factor of roughly 2.}
 \label{fnlconstraints}
\end{figure}

Figure \ref{fnlconstraints} shows how the constraints on the square root of the power spectrum change depending on the value of $f_{NL}$. For positive values of $f_{NL}$ we see that the constraints tighten (corresponding to an increase in the abundance of PBHs for a given value for the power spectrum, see Figure \ref{fnlpdfs}). For negative values, we see that the constraints weaken dramatically - this is because, unless $\sigma$ becomes large, no PBHs form at all. As $f_{NL}$ becomes significantly negative, we see that the constraints for $\beta=10^{-20}$ and $\beta=10^{-5}$ converge. Unless there is remarkable fine tuning in the size of the perturbations at small scales, there would either be far too many PBHs, or none. Using this method to calculate the constraints, as $f_{NL}$ becomes more negative the constraints on the power spectrum do flatten out at a value above 1 - however, the perturbative approach does not work when the perturbation amplitude is ${\cal O}(1)$ or higher, so these results cannot be trusted.

\subsection{Cubic non-Gaussianity}

The model of local non-Gaussianity is now taken to be
\begin{equation}
\label{cubicNG}
\zeta=\zeta_{g}+\frac{9}{25}g_{NL}\zeta_{g}^{3}.
\end{equation}
We follow the same process as before to calculate the PDFs and constraints on the power spectrum \cite{AK}. Care needs to be taken with the amount of solutions to Eq.~(\ref{cubicNG}). For $g_{NL}>0$, there is one solution for all $\zeta$. But for $g_{NL}<0$, there may be multiple solutions. For example, for $g_{NL}<0$, in the range
\begin{equation}
-\frac{2}{9}\sqrt{\frac{-5}{g_{NL}}}\leq\zeta\leq\frac{2}{9}\sqrt{\frac{-5}{g_{NL}}},
\end{equation}
there are 3 solutions to Eq.~\ref{cubicNG}. These solutions need to be taken into account when calculating PDFs or constraints on the power spectrum.

\begin{figure}[t]
\centering
\begin{subfigure}{0.5\textwidth}
	\centering
	\includegraphics[width=\linewidth]{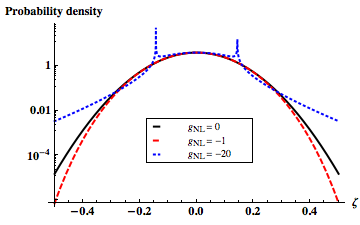}
\end{subfigure}%
\begin{subfigure}{0.5\textwidth}
	\centering
	\includegraphics[width=\linewidth]{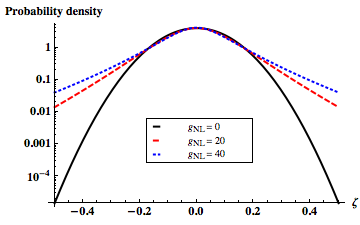}
\end{subfigure}
\begin{subfigure}{0.5\textwidth}
	\centering
	\includegraphics[width=\linewidth]{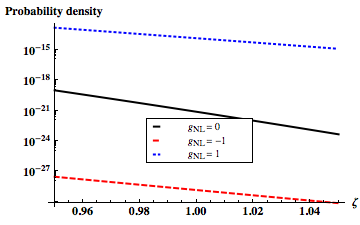}
\end{subfigure}
 \caption{The top left (right) plot shows the effect of negative (positive) $g_{NL}$ on the PDF. We see that $g_{NL}$ affects the kurtosis of the distribution. Positive $g_{NL}$ always gives a sharper peak with broader tails - enhancing PBH production. Large negative $g_{NL}$ has a similar effect - however, we see two sharp peaks in the distribution, due to the derivative in Eq.~(\ref{changeofvariable}) becoming infinite. For small negative $g_{NL}$ we see that the tails of the distribution are diminished. The bottom plot shows the tail of the PDF where PBHs form - again showing a very strong dependence on small amounts of non-Gaussianity, and again the sign of the non-Gaussianity is important. For these plots, we have again taken $\sigma=0.1$.}
 \label{gnlpdfs}
\end{figure}

Figure \ref{gnlpdfs} shows a log plot of the effects of $g_{NL}$ on the PDF. The upper left (right) panel shows the effect of positive (negative) $g_{NL}$. We see that $g_{NL}$ affects the kurtosis of the distribution. Typically, serving to give a distribution which is more sharply peaked in the central region, but with larger tails. Positive $g_{NL}$ always serves to enhance the amplitude of the tails where PBHs form, as does large negative $g_{NL}$. However, for small negative $g_{NL}$ the tails of the PDF are diminished - leading to a lower PBH abundance (and consequently, weaker constraints). 

\begin{figure}[t]
\centerline{ \includegraphics[scale=0.8]{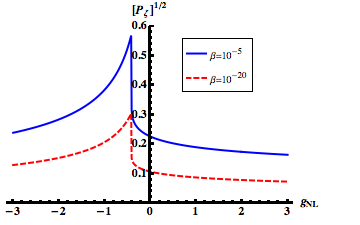}}
 \caption{This plot shows how the constraints on the square root of the power spectrum due to PBHs depend on $g_{NL}$. }
\label{gnlconstraints}
\end{figure}

In order to calculate the constraints on the power spectrum, we again write an expression for $\beta$ to be solved. For positive $g_{NL}$ we have
\begin{equation}
\beta\simeq\frac{1}{\sqrt{2\pi}}\int_{y_{1}}^{\infty}e^{-\frac{y^2}{2}}dy.
\end{equation}
For $\frac{-20}{81}<g_{NL}<0$, there are 3 solutions to Eq.~(\ref{cubicNG}), and $\beta$ is given by
\begin{equation}
\label{betagnl}
\beta\simeq\frac{1}{\sqrt{2\pi}}\left(\int_{-\infty}^{y_{1}}e^{-\frac{y^2}{2}}dy+\int_{y_{2}}^{y_{3}}e^{-\frac{y^2}{2}}dy\right).
\end{equation}
Finally, for $g_{NL}<-\frac{20}{81}$, $\beta$ is given by
\begin{equation}
\beta\simeq\frac{1}{\sqrt{2\pi}}\int_{-\infty}^{y_{1}}e^{-\frac{y^2}{2}}dy.
\end{equation}
The limits on the integrals here ($y_{1}$, $y_{2}$, etc) are solutions for $y$ to Eq.~(\ref{cubicNG}). The variance in this model is given by \cite{Y}
\begin{equation}
{\cal P}_{\zeta}=\sigma^2\left(1+\frac{54}{25}g_{NL}\sigma^{2}\ln(kL)+27\left(\frac{9g_{NL}}{25}\right)^{2}\sigma^{4}\ln(kL)^{2}\right).
\end{equation}
Figure \ref{gnlconstraints} shows the constraints obtained for the cubic non-Gaussianity model. For small $g_{NL}$ we see that the constraints on the power spectrum are highly asymmetric between positive and negative $g_{NL}$. This is because for positive $g_{NL}$ an overdensity in the linear $\zeta$ regime is boosted by the cubic term - especially so in the tail of the PDF, and so the constraints tighten. However, for small negative $g_{NL}$ the opposite is the case and the two terms tend to cancel each other, and hence the constraints weaken dramatically. For very small negative $g_{NL}$, the 2nd term in the expression for $\beta$, Eq.~(\ref{betagnl}), dominates. As $g_{NL}\rightarrow-\frac{20}{81}$ from above, $y_{3}-y_{2}\rightarrow0$, and this term decreases rapidly so that the constraint on the power spectrum rapidly becomes weaker. As $g_{NL}$ becomes more negative, the first term in Eq.~(\ref{betagnl}) increases, and the constraints tighten again. As $g_{NL}$ becomes large, either positive or negative, then the cubic term in Eq.~(\ref{cubicNG}) dominates the expression, $\zeta\propto\pm\zeta_{g}^{3}$, and the constraints don't depend on the sign of $g_{NL}$. This is because the Gaussian PDF is invariant under a change of sign of $\zeta_{g}$, which is equivalent to changing the sign of $g_{NL}$ (in the case where the linear term is absent). For this reason, the constraints asymptote to the same value as $|g_{NL}|\rightarrow\infty$.

\subsection{Higher order terms}\label{sec:higher-order}

In this section, we consider the effects of higher order terms on the constraints that can be placed on the power spectrum. We take the model of local non-Gaussianity to be
\begin{equation}
\label{fullexpansion}
\zeta=\zeta_{g}+\frac{3}{5}f_{NL}\left(\zeta_{g}^{2}-\sigma^{2}\right)+\frac{9}{25}g_{NL}\zeta_{g}^3+\frac{27}{125}h_{NL}\left(\zeta_{g}^{4}-3\sigma^{4}\right)+\frac{81}{625}i_{NL}\zeta_{g}^5+\cdots .
\end{equation}
Higher order terms have a similar effect on the PDF as do the quadratic and cubic terms - even order terms introduce skew-like asymmetry to the PDF, whilst odd order terms affect kurtosis, and have similar effects on the tails of the pdfs.

\begin{figure}[t]
\centerline{ \includegraphics[scale=0.8]{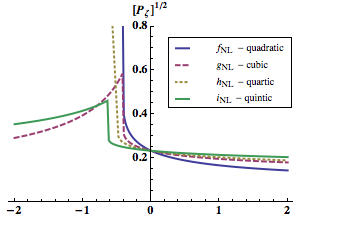}}
 \caption{Here we see how the constraints on the square root of the power spectrum depends on non-Gaussianity to $5^{th}$ order. We have considered the addition of each order term one at a time. Note that the even order terms display similar behaviour to each other, as do the odd order terms. The constraints here are shown for the case $\beta=10^{-5}$. Here, we have included only the linear term and one other term in Eq.~(\ref{fullexpansion}) for each order equation. The x-axis is either $f_{NL}$, $g_{NL}$, $h_{NL}$, or $i_{NL}$, depending on the order equation being used.}
\label{fghi}
\end{figure}

The number of solutions to $\zeta(\zeta_g)=1$ depends on the values of $f_{NL}$, $g_{NL}$, $h_{NL}$, etc. Because an analytic solution is not typically available for polynomial equations above 4th order, a numerical method was used to calculate the constraints on the power spectrum. Starting from the linear, purely Gaussian model, a value for $\sigma$ is calculated. The non-Gaussianity parameters are then varied slowly, and Eq.~(\ref{fullexpansion}) is solved using the previous value of $\sigma$ to find critical values of $\zeta_{g}$ required for PBH formation,
\begin{equation}
\zeta_{g}(\zeta_{c})=\zeta_{g1}, \zeta_{g2},\cdots .
\end{equation}
As before, a Gaussian variable $y$ with unit variance is used, Eq.~(\ref{y}), and an expression for $\beta$ is written. For example,
\begin{equation}
\beta\simeq\frac{1}{\sqrt{2\pi}}\left(\int_{y_{1}}^{y_{2}}e^{-\frac{y^2}{2}}dy+\int_{y_{3}}^{y_{4}}e^{-\frac{y^2}{2}}dy+...\right).
\end{equation}
This is then solved numerically to find a value for $\sigma$ and the variance is calculated. Provided that small enough steps are taken, and that $\sigma$ varies sufficiently slowly, the results obtained through this method are in excellent agreement to those obtained previously by an analytic method. Accounting for terms to $5^{th}$ order in $\zeta$ and including all orders in loops, using the techniques of \cite{Y} we find that the power spectrum is given by
%
%
\begin{eqnarray}
{\cal P}_{\zeta}&=&\sigma^{2}+\left(\frac{3}{5}\right)^2 \left( 4 f_{NL}^2+6 g_{NL}\right)\sigma^{4}\ln(kL)+ \left(\frac{3}{5}\right)^4 \left( 27 g_{NL}^2+ 48 f_{N}h_{NL}+30 i_{NL}\right) \sigma^{6}\ln(kL)^2 \nonumber \\
&&+\left(\frac{3}{5}\right)^6 \left( 240 h_{NL}^2+450 g_{NL} i_{NL}\right)\sigma^{8}\ln(kL)^3 +
\left(\frac{3}{5}\right)^{8}  2625 i_{NL}^2\sigma^{10}\ln(kL)^4.
\end{eqnarray}
Figure \ref{fghi} shows how the constraints on the power spectrum depend upon the non-Gaussianity parameters. Here, we consider the effects of each term in Eq.~(\ref{fullexpansion}) one at a time. Again, for higher order terms, we see similar behaviour to that seen for the quadratic and cubic non-Gaussianity. For even-order terms, the constraints become tighter for positive values, but weaken dramatically even for small negative values. For odd-order terms, the constraints become tighter for positive values, but for small negative values, the constraints initially weaken dramatically before tightening again. The constraints are most sensitive to small negative non-Gaussianity - where the positive tail of the PDF is strongly reduced, either due to a skew-like asymmetry in the PDF from even terms, or kurtosis type effects from the odd terms. 

\section{Hierarchical scaling}

In order to study the effects of the different types of local non-Gaussianity simultaneously, we introduce some hierarchical scaling relationship between the non-Gaussianity parameters. Here, we present the simple idea of a power law scaling between the terms:
\begin{equation}
\label{powerlaw}
g_{NL}\sim \alpha^{2} f_{NL}^{2},
h_{NL}\sim \alpha^{3} f_{NL}^{3},
i_{NL}\sim \alpha^{4} f_{NL}^{4}, \cdots,
\end{equation}
where $\alpha$ is a constant of order unity, and the model of local non-Gaussianity can be taken as
\begin{equation}
\zeta\sim\zeta_{g}+\frac{3}{5}f_{NL}\left(\zeta_{g}^{2}-\sigma^{2}\right)+\frac{9}{25}\alpha^{2}f_{NL}^{2}\zeta_{g}^3+\frac{27}{125}\alpha^{3}f_{NL}^{3}\left(\zeta_{g}^{4}-3\sigma^{4}\right)+\frac{81}{625}\alpha^{4}f_{NL}^{4}\zeta_{g}^5+\cdots.
\end{equation}
This type of relation can occur in several different models, including multi-brid inflation \cite{AA,AB}, a similar scaling was used in \cite{AF}.

Figure \ref{powerlaw} shows the effect of the hierarchical scaling to the constraints on the power spectrum to different orders, where we have taken $\alpha=1$ (modifying this term but keeping it of order unity does not significantly affect the results). When calculating to $n^{th}$ order, we have now included all terms up to and including the $n^{th}$ term (rather than just the single term in the previous section). Again, we see similar behaviour for the different order expansions - depending on whether the highest order term is even or odd.

For positive $f_{NL}$ the constraints tighten significantly as $f_{NL}$ increases, before converging to some constant as $f_{NL}\rightarrow\infty$. As $f_{NL}$ becomes large however, the highest-order term dominates Eq.~(\ref{fullexpansion}), and it is sufficient to take, for example, $\zeta\propto\zeta_{g}^{n}$. Note that the constraints found in this region depend on the order that Eq.~(\ref{fullexpansion}) is taken to - the constraints are slightly tighter for higher orders.

For negative $f_{NL}$, we see similar behaviour to that seen before when only a single term was considered. When the highest order term is even the constraints weaken dramatically as $f_{NL}$ becomes negative,  again requiring fine tuning to produce any PBHs without overproducing them. When the highest order terms are odd, we again see a peak where the constraints weaken for small negative values, before slowly tightening - however, the peak is now smoother. Again, as $|f_{NL}|\rightarrow\infty$ and for odd terms, the sign of the non-Gaussianity parameter does not matter, and the constraints approach the same value. Whilst this may not be obvious from Fig. \ref{powerlaw}, if the axes were extended to large $f_{NL}$, of order $10^{4}$, we would see this to be the case.

\begin{figure}[t]
\centerline{ \includegraphics[scale=0.8]{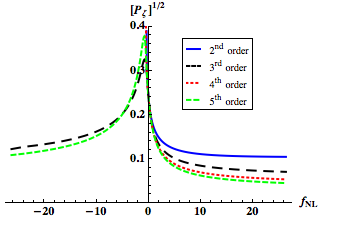}}
 \caption{The constraints on the square root of the power spectrum in the case of a hierarchical power law between the non-Gaussianity parameters. The constraints here are shown for the case $\beta=10^{-5}$. We have used here the hierarchical power rule to different orders (up to $5^{th}$ order), and show the constraints obtained in each case change depending on $f_{NL}$. Note that we see two distinct behaviours - depending on whether the highest term on the expansion is odd or even - which give very different results for the case of negative $f_{NL}$.}
\label{powerlaw}
\end{figure}

\section{PBHs in the curvaton model}

Whilst the simplest inflationary models give rise to a nearly Gaussian distribution of the primordial curvature perturbation, multi-field models of inflation can lead to strong non-Gaussianity. One well motivated model is the curvaton model \cite{AD}. In this model, in addition to the field driving inflation, the inflaton $\phi$, there is a second light scalar field, the curvaton $\chi$, whose energy density is completely subdominant during inflation. At Hubble exit during inflation both fields acquire classical perturbations that freeze in. Here, the observed perturbations in the CMB and LSS, as well as perturbations on smaller scales, can result from the curvaton instead of the inflaton. At the end of inflation, the inflaton decays into relativistic particles (``radiation"). The curvaton energy density is still sub-dominant at this stage and carries an isocurvature perturbation - and at some later time, the curvaton also decays into radiation. Taking the curvaton to be non-relativistic before it decays, the energy density of the curvaton will decay slower than the energy density of the background radiation - and consequently the curvature perturbation due to the curvaton will become dominant. 

If the curvaton generates the perturbations on CMB scales, then in simple realisations of the curvaton scenario with a quadratic potential it cannot have a much larger amplitude of perturbations on smaller scales. However it is possible that a second stage of inflation has a dominant contribution to its perturbations from the curvaton model. Indeed if the curvaton mass, $m_{\chi}$, is reasonably heavy compared to the Hubble scale, then it will naturally have a blue spectrum giving the smallest scale perturbations the largest amplitude. The spectral index is given by $n_s-1=2 m^2_{\chi}/(3H^2)+2\dot{H}/H^2$, where all quantities should be evaluated at the horizon crossing time of the relevant scale \cite{AD}. Motivated by our discovery in the last section that truncating the pdf to any order in the non-linearity parameters can give a very bad approximation to the true result, a practical reason for studying the curvaton scenario is that this is a rare case in which the full non-linear pdf has been calculated. This allows us to check in a realistic and popular model whether the non-Gaussian corrections to the pdf are important, and whether just including the first few terms such as $f_{NL}$ or $g_{NL}$ would give an accurate result. We will see that the non-Gaussian corrections to all orders are always important when studying PBH formation.

\begin{figure}[t]
\centerline{ \includegraphics[scale=0.8]{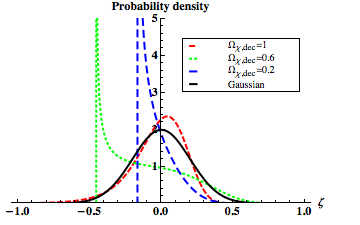}}
 \caption{PDFs in the curvaton model. Here we see that, whilst $\Omega_{\chi,dec}\sim1$ the distribution is close to Gaussian. However, as $\Omega_{\chi,dec}$ the PDF becomes more non-Gaussian, enhancing the positive tail of the PDF. These have been calculated using a formal change of variable using \ref{curvatonequation} and \ref{nononlinear}. All the plots have a variance $\langle \zeta^{2}\rangle=0.04$.}
\label{curvatonpdfs}
\end{figure}

Here we use the result obtained by Sasaki et al in the \emph{sudden decay approximation} \cite{AD}:
\begin{equation}
\label{curvrad}
\left(1-\Omega_{\chi,dec}\right)e^{4(\zeta_{r}-\zeta)}+\Omega_{\chi,dec}e^{3(\zeta_{\chi}-\zeta)}=1,
\end{equation}
where $\Omega_{\chi,dec}$ is the dimensionless curvaton density parameter for the curvaton at the decay time. Taking the curvature perturbation in the radiation fluid to be negligible, i.e. $\zeta_{r}=0$, Eq.~(\ref{curvrad}) reads
\begin{equation}
\label{curvatonequation}
e^{3\zeta_{\chi}}=\frac{1}{\Omega_{\chi,dec}}\left(e^{3\zeta}+(\Omega_{\chi,dec}-1)e^{-\zeta}\right).
\end{equation}
This gives the fully non-linear relation between the primordial curvature perturbation, $\zeta$, and the curvaton curvature perturbation, $\zeta_{\chi}$. Taking there to be no non-linear evolution between Hubble exit and the start of curvaton decay, the left hand side of Eq.~(\ref{curvatonequation}) is given by
\begin{equation}
\label{nononlinear}
e^{3\zeta_{\chi}}=\left(1+\frac{\delta_{1}\chi}{\bar{\chi}}\right)^{2},
\end{equation}
where $\delta_{1}\chi$ is the Gaussian perturbation in the curvaton field at Hubble exit, and $\bar{\chi}$ is the background value. Eq.~(\ref{curvatonequation}) is quartic in $e^{\zeta}$ and so this allows us to write an expression for the full curvature perturbation, $\zeta$, in terms of the Gaussian variable $\delta_{\chi}=\frac{\delta_{1}\chi}{\bar{\chi}}$, or equivalently write the Gaussian variable as a function of the curvature perturbation.
\begin{equation}
\delta_{\chi}=\delta_{\chi}(\zeta).
\end{equation}
Note that, for $\Omega_{\chi,dec}<1$, $\zeta$ is bounded from below, with the minimum value given by
\begin{equation}
\zeta_{min}=\frac{1}{4}\ln\left(1-\Omega_{\chi,dec}\right).
\end{equation}
Making a formal change of variable allows the PDF to be calculated. Figure \ref{curvatonpdfs} shows the PDFs obtained for different values of $\Omega_{\chi,dec}$. Whilst $\Omega_{\chi,dec}$ is close to unity, the PDF is close to Gaussian - however, the positive tail of the PDF is diminished, reducing PBH formation. As $\Omega_{\chi,dec}$ becomes smaller, the PDF becomes more strongly non-Gaussian, and the positive tail of the PDF is enhanced, increasing PBH formation.

Constraints on the power spectrum are obtained using the same method as before. Eq.~(\ref{curvatonequation}) is solved for $\zeta=\zeta_{c}$ to find the corresponding critical values of $\delta_{\chi}$, giving two solutions, $\delta_{c1}$ and $\delta_{c2}$, for all values of $\Omega_{\chi,dec}$. An expression for $\beta$ is written:
\begin{equation}
\beta\simeq\frac{1}{\sqrt{2\pi}\sigma}\left(\int^{\infty}_{\delta_{c1}}e^{-\frac{\delta_{\chi}^{2}}{2\sigma}}d\delta_{\chi}+\int^{\delta_{c2}}_{-\infty}e^{-\frac{\delta_{\chi}^{2}}{2\sigma}}d\delta_{\chi}\right).
\end{equation}
This expression is then solved numerically to find a value for $\sigma$ for a given value of $\beta$. Now that all of the necessary components have been found, the constraints on the power spectrum are calculated by finding the variance through numeric integration
\begin{equation}
{\cal P}_{\zeta}=\int_{\zeta_{min}}^{\infty}\zeta^{2}P_{NG}(\zeta)d\zeta=\int_{-\infty}^{\infty}\zeta(\chi_{g})^{2}P_{G}(\chi_{g})d\chi_{g},
\end{equation}
where $P_{NG}(\zeta)$ and $P_{G}(\chi_{g})$ are the non-Gaussian and Gaussian PDF's respectively. Care needs to be taken to ensure that the mean of $\zeta$ is zero during the calculation - if necessary defining a new variable with the mean subtracted, such that $\langle\zeta\rangle=0$.

\begin{figure}[t]
\centerline{ \includegraphics[scale=0.8]{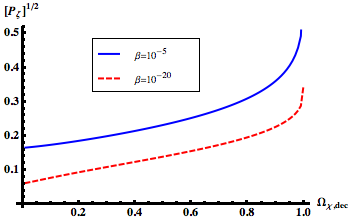}}
 \caption{Constraints on the square root of the power spectrum in the curvaton model. The constraints obtained for different constraints on $\beta$, the initial PBH mass fraction, as a function of $\Omega_{\chi,dec}$, the dimensionless curvaton density parameter at the time of decay.}
\label{curvatonconstraints}
\end{figure}

Figure \ref{curvatonconstraints} shows the constraints obtained for different values of $\beta$. When $\Omega_{\chi,dec}\sim1$, the constraints are weaker than in the Gaussian case even though the PDF is close to Gaussian - this is an example of even small amounts of non-Gaussianity having a large impact on the constraints. As $\Omega_{\chi,dec}\rightarrow0$, the constraints on the power spectrum become tighter, corresponding to an enhancement of the positive tail of the PDF.

\begin{figure}[t]
\centerline{ \includegraphics[scale=0.8]{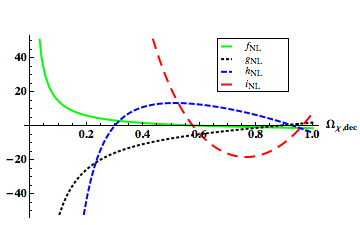}}
 \caption{The non-Gaussianity parameters in the curvaton model.}
\label{curvatonparameters}
\end{figure}

It should be noted that, in this model, a full expansion for $\zeta$ can be obtained by performing a Taylor expansion of the solution to Eq.~(\ref{curvatonequation}) \cite{AD,AA}. Figure \ref{curvatonparameters} shows the non-Gaussianity parameters plotted as a function of $\Omega_{\chi,dec}$. Instead of using the full non-linear solution for $\zeta$, the calculation can be completed as in the previous section by using these solutions for the parameters. However, the results obtained in this manner typically do not match well with those obtained from an analytic solution - the contributions to the power spectrum from higher-order terms can become large and can be either positive or negative. This is due to the fact that, whatever order the expansion is carried out to, the Taylor expansion diverges from the analytic solution as $\zeta$ becomes large (of order unity or higher). For example, for $\Omega_{\chi,dec}=1$, $f_{NL}=-\frac{5}{4}$, and so a truncation at second order would not even come close to matching with the results obtained here. Comparing the constraints for $\beta=10^{-5}$ between Figs.~\ref{curvatonconstraints} and \ref{fnlconstraints}, notice that the Gaussian constraint of ${\cal P}_{\zeta}^{1/2}=0.23$ is reached for $\Omega_{\chi,dec}\simeq0.4$, but from Fig. \ref{curvatonparameters} we see that the non-linearity parameters are not typically small here, and so the matching is just coincidence. Hence we conclude that the non-Gaussianity of the curvaton model always has to be taken into account when calculating PBH constraints. 

\section{Summary}

The formation rate of PBHs probes the tails of the PDF of primordial fluctuations, and is very sensitive to the effects of non-Gaussianity. We have calculated the effects of the local model of non-Gaussianity for terms up to 5th order, parameterised by $f_{NL}$, $g_{NL}$, $h_{NL}$, and $i_{NL}$. We have shown that any non-Gaussianity parameters of order unity can have a significant effect on the abundance of PBHs, and the constraints that can be placed on the power spectrum - due to the fact that the non-Gaussianity parameters have a large impact on the tails of the PDF.

The sign of the non-Gaussianity has a particularly strong effect. We see that positive terms of even order tighten the constraints significantly, but negative terms dramatically weaken the constraints, to the point where the curvature perturbation is order unity. Typically, when an even type of non-Gaussianity is considered, such as $f_{NL}$ or $h_{NL}$, if this term is negative and dominates the non-Gaussianity of the distribution, the amplitude of the primordial fluctuations will either be too small to form any PBHs, or so large that the Universe contains too many PBHs. Such a scenario would be incompatible with any future detection of PBHs. Odd-order terms, such as $g_{NL}$ or $i_{NL}$, tend to tighten the constraints regardless of their sign, but small negative terms can weaken the constraints dramatically over a small range of values. If PBHs were to be detected in the future, they could potentially rule out certain models and distributions. Care needs to be taken as truncations to set order in the model of non-Gaussianity used might not converge.

In the curvaton model, the PDF is relatively close to Gaussian if the Universe is dominated by the curvaton at the time of decay, $\Omega_{\chi,dec}\sim1$ - and in this case the constraints are weakened compared to a purely Gaussian distribution. As $\Omega_{\chi,dec}$ decreases, the distribution becomes more non-Gaussian, and the constraints on the power spectrum tighten significantly. Calculations obtained for the curvaton model by calculating the local non-Gaussianity parameters to e.g.~second or third order ($f_{NL}$ or $g_{NL}$) do not agree with those obtained using the full non-linear solution. Therefore, given a specific model, it may be necessary to calculate the full hierarchy (rather than truncating at a given order) before performing calculations, as we have done here for the curvaton model.

\section{Acknowledgements}
We would like to thank Anne Green and David Seery for useful discussions. SY thanks the STFC for financial support. CB is supported by a Royal Society University Research Fellowship.

\bibliographystyle{JHEP}
\bibliography{bibdeskfile.bib}

\end{document}